\documentclass[aps,prl,twocolumn,showpacs,superscriptaddress]{revtex4-1}
\usepackage{graphicx}

\begin{document}

\title{Interface-driven ferromagnetism within the quantum wells of a rare earth titanate superlattice}

\author{R. F. Need}
 \affiliation{Materials Department, University of California, Santa Barbara, California 93106-5050, USA}
 
\author{B. J. Isaac}
 \affiliation{Materials Department, University of California, Santa Barbara, California 93106-5050, USA}

\author{B. J. Kirby}
 \affiliation{NIST Center for Neutron Research, National Institute of Standards and Technology, Gaithersburg, Maryland, USA}

\author{J. A. Borchers}
 \affiliation{NIST Center for Neutron Research, National Institute of Standards and Technology, Gaithersburg, Maryland, USA}

\author{S. Stemmer}
 \affiliation{Materials Department, University of California, Santa Barbara, California 93106-5050, USA}

\author{Stephen D. Wilson}
 \email{stephendwilson@engineering.ucsb.edu}
 \affiliation{Materials Department, University of California, Santa Barbara, California 93106-5050, USA}
 
\begin{abstract}
Here we present polarized neutron reflectometry measurements exploring thin film heterostructures comprised of a strongly correlated Mott state, GdTiO$_3$, embedded with SrTiO$_3$ quantum wells.  Our results reveal that the net ferromagnetism inherent to the Mott GdTiO$_3$ matrix propagates into the nominally nonmagnetic SrTiO$_3$ quantum wells and tracks the magnetic order parameter of the host Mott insulating matrix.  Beyond a well thickness of 5 SrO layers, the magnetic moment within the wells is dramatically suppressed, suggesting that quenched well magnetism comprises the likely origin of quantum critical magnetotransport in this thin film architecture.  Our data demonstrate that the interplay between proximate exchange fields and polarity induced carrier densities can stabilize extended magnetic states within SrTiO$_3$ quantum wells.
\end{abstract}

\maketitle
Interfaces between RETiO$_3$ (RE=Gd, Sm, $...$) and SrTiO$_3$ aggregate charge via a polar discontinuity between the differing valence states intrinsic to the rare earth and alkali earth layers of the two compounds \cite{6}.  In multilayer films, two sequential interfaces define a quantum well into which the polarization-induced carriers preferentially spread \cite{6,7}. This induced charge lives in a physically rich landscape; one where traversing between sufficiently thick layers also necessitates the relaxation of \textit{d}-electron orbital polarization, long-range magnetic order, and strong on-site Coulomb interactions \cite{8}.  The thickness of the quantum well also defines a length scale for the mediation of interactions between polarity-induced carriers, which at sufficiently high densities have the potential to drive electronic order \cite{9}.  Adding further complexity, structural symmetries (i.e. oxygen octahedral tilts) from the parent RETiO$_3$ can coherently propagate across the interface and into the well before relaxing beyond a critical thickness \cite{10,11,12}.  Ultimately, the combination of these effects may modify the bandwidth and electronic states manifest within the well, generating a parameter space not realizable in bulk form.

The interplay between polar interface charge and a proximate correlated state renders exotic transport phenomena in SmTiO$_3$/SrTiO$_3$ and GdTiO$_3$/SrTiO$_3$ heterostructures \cite{13,14}.  The band insulator SrTiO$_3$, when embedded as thin quantum wells within Mott insulating GdTiO$_3$ barriers, exhibits metallic transport mediated via interface carriers \cite{15}.  A metal to insulator transition (MIT) emerges as the well thickness (defined by the number of SrO layers) decreases to 2 SrO layers and the corresponding well carrier density diverges \cite{18}.  Prior to this MIT, SrTiO$_3$ quantum wells with thicknesses of approximately 3 SrO layers display an unusual hysteresis in their low temperature longitudinal magnetoresistance---a state suggestive of domain switching and a field coupled electronic order parameter \cite{16,17}.  

Intriguingly, a divergent carrier mass was also observed near the stabilization of this order parameter, consistent with a quantum critical point (QCP) driven by the well carrier density/dimensionality \cite{18}.  However, little remains understood regarding the origins of this unusual phase behavior in GdTiO$_3$/SrTiO$_3$ heterostructures absent a direct resolution of the order parameter within the wells.  Addressing this and searching for the presence of interface-induced magnetic order requires access to an experimental probe sensitive to magnetic polarization and capable of resolving its depth profile throughout a heterostructure---both of which are achievable via polarized neutron reflectometry (PNR) \cite{19,20,21}.  

In this paper, we present a PNR study exploring magnetic order within the quantum wells of GdTiO$_3$/SrTiO$_3$ heterostructures.  Our data reveal the presence of magnetization induced within the SrTiO$_3$ wells below a critical well thickness of 5 SrO and demonstrate a novel realization of magnetic order induced within a nonmagnetic medium through the interplay between polarity induced charge density and proximity induced magnetic exchange.  Furthermore, our results suggest that well magnetism represents the local order parameter whose suppression generates the divergent carrier mass reported in earlier magnetotransport studies \cite{18}.   

A series of superlattice films containing a quintuple of 4 nm GdTiO$_3$ spacer layers separated by variable width SrTiO$_3$ quantum wells (2 SrO, 3 SrO, 5 SrO, and 10 SrO layers) were grown via molecular beam epitaxy, and PNR measurements were collected at the NIST Center for Neutron Research on the PBR reflectometer.  PNR models of magnetism in GdTiO$_3$/SrTiO$_3$ heterostructures benefit from independent measurements of the films' structures in order to constrain the number of free-parameters. To achieve this, structural profiles were collected via room temperature x-ray reflectivity (XRR) measurements \cite{Supplemental}, and the XRR layer thicknesses and effective roughnesses were used as a fixed input in subsequent PNR models.  Further details regarding film growth and reflectometry experiments are provided in supplementary materials \cite{Supplemental} and detailed electron microscopy characterization of the film interfaces are described elsewhere \cite{7,10,15}.  While transmission electron microscopy (TEM) measurements of identical superlattice films show sharp interfaces \cite{7,10,15} and a maximum chemical intermixing of one atomic layer \cite{15}, roughnesses measured by reflectivity are an average across the entire sample surface area, including effects from step edges across the underlying substrate.   Hence, modeling quantum wells in the thin layer limit renders widths, roughnesses, and scattering length densities (SLDs) whose values become intrinsically coupled.  As such, the refined roughness values in the thin well limit do not have an independent physical meaning and should not be compared to local probe measurements (e.g. TEM).   

\begin{figure}
\includegraphics[scale=0.375]{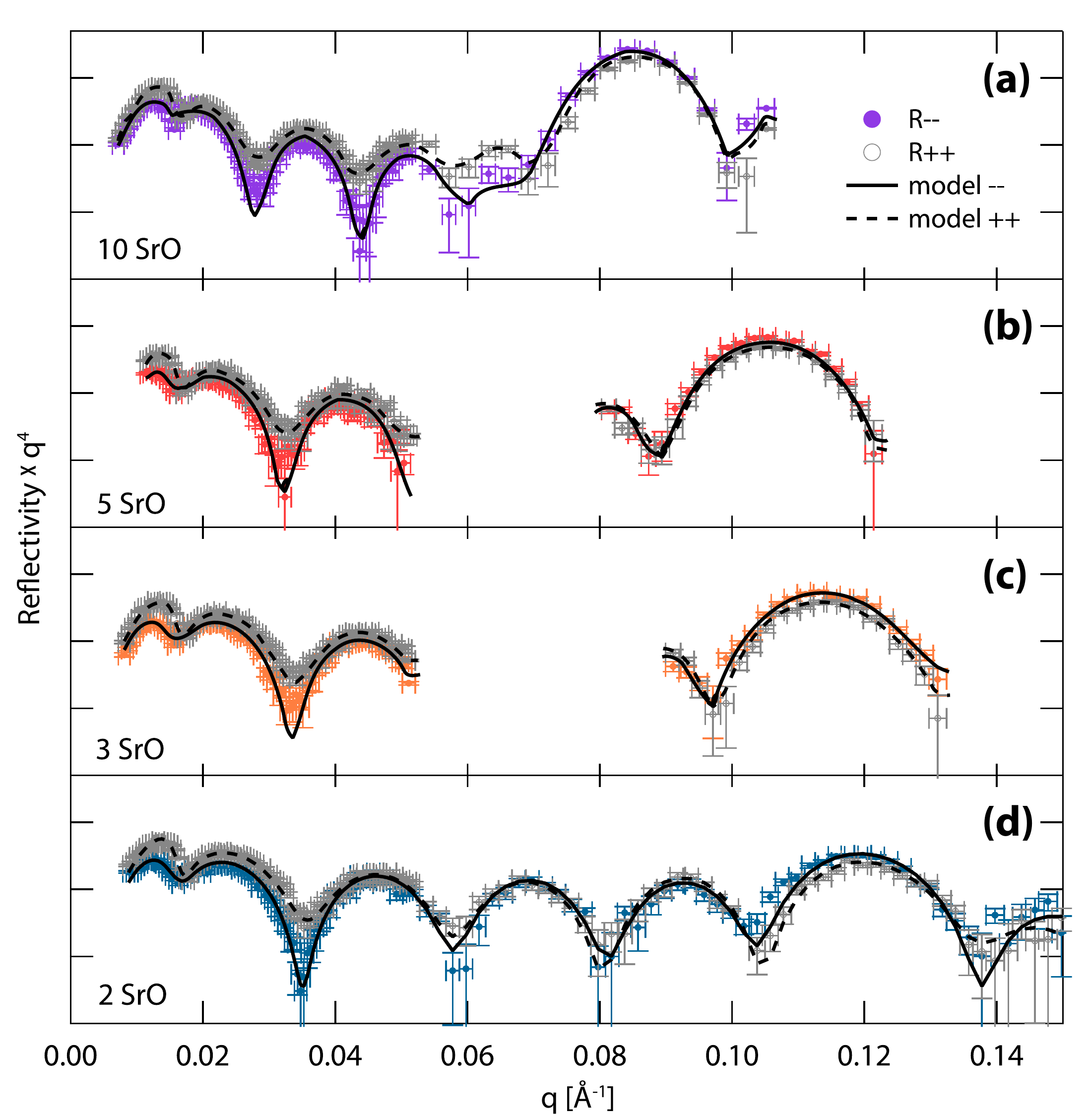}
\caption{PNR data of GdTiO$_3$/SrTiO$_3$ superlattice films.  (a)-(d) Neutron reflectivity as a function of momentum transfer for GdTiO$_3$/SrTiO$_3$/GdTiO$_3$ superlattice structures with quantum wells of thickness 10 SrO, 5 SrO, 3SrO, and 2 SrO layers respectively.  Open gray symbols denote the $R^{++}$ cross section and closed color symbols denote $R^{--}$.  Dashed lines ($R^{++}$) and solid lines ($R^{--}$) indicate the fits resulting from structural and magnetic models of the data. }  
\end{figure}

Figure 1 shows the results from low temperature ($T = 4$ K) PNR measurements on four superlattice samples collected under a field-cooled (FC) state ($\mu_0 H = 0.7$ T).  Solid and empty symbols show non-spin flip data collected for specular reflectivity curves with the incident and scattered neutron polarizations oriented down ($R^{--}$) and up ($R^{++}$) relative to the sample field respectively, both of which encode information regarding the nuclear and magnetic SLD profiles of the film \cite{28,29}.  While the overall oscillation of both curves is primarily reflective of the chemical profile of the film, splitting between these curves denotes a net, in-plane, magnetic polarization along the field direction where changes in magnetization between layers produce a difference in the ($R^{++} - R^{--}$) cross section.  Chemical and magnetic scattering profiles can be modeled simultaneously via an optical matrix formalism \cite{30,Supplemental}, and the resulting fits are plotted in Fig. 1.  Gaps in the data sets (i.e. $q = 0.055-0.080$ $\AA^{-1}$ in Figs. 2 (b) and (b)) are due to limited measurement time and prioritization of $q$ ranges where SrTiO$_3$ features are most salient.  

The parameters summarizing the modeled films' depth profiles at 4 K and 30 K are plotted in Fig. 2.  Chemical contrast varies as expected between GdTiO$_3$ and SrTiO$_3$ layers with the topmost GdTiO$_3$ layer distinct from the buried layers due to brief periods of exposure to atmosphere.  The effective roughnesses of the GdTiO$_3$$\rightarrow$SrTiO$_3$ and SrTiO$_3$$\rightarrow$GdTiO$_3$ interfaces span the thicknesses of the wells in the 2 SrO and 3 SrO samples; however the wells in the 5 SrO and 10 SrO samples are able to reach their nominal bulk SLDs and increasingly decouple from neighboring layers.  To better isolate magnetism in the wells, nuclear SLDs were refined at low temperature and then fixed \cite{Supplemental}; only the magnetic neutron SLDs were allowed to vary as a function of temperature. 

\begin{figure}
\includegraphics[scale=0.375]{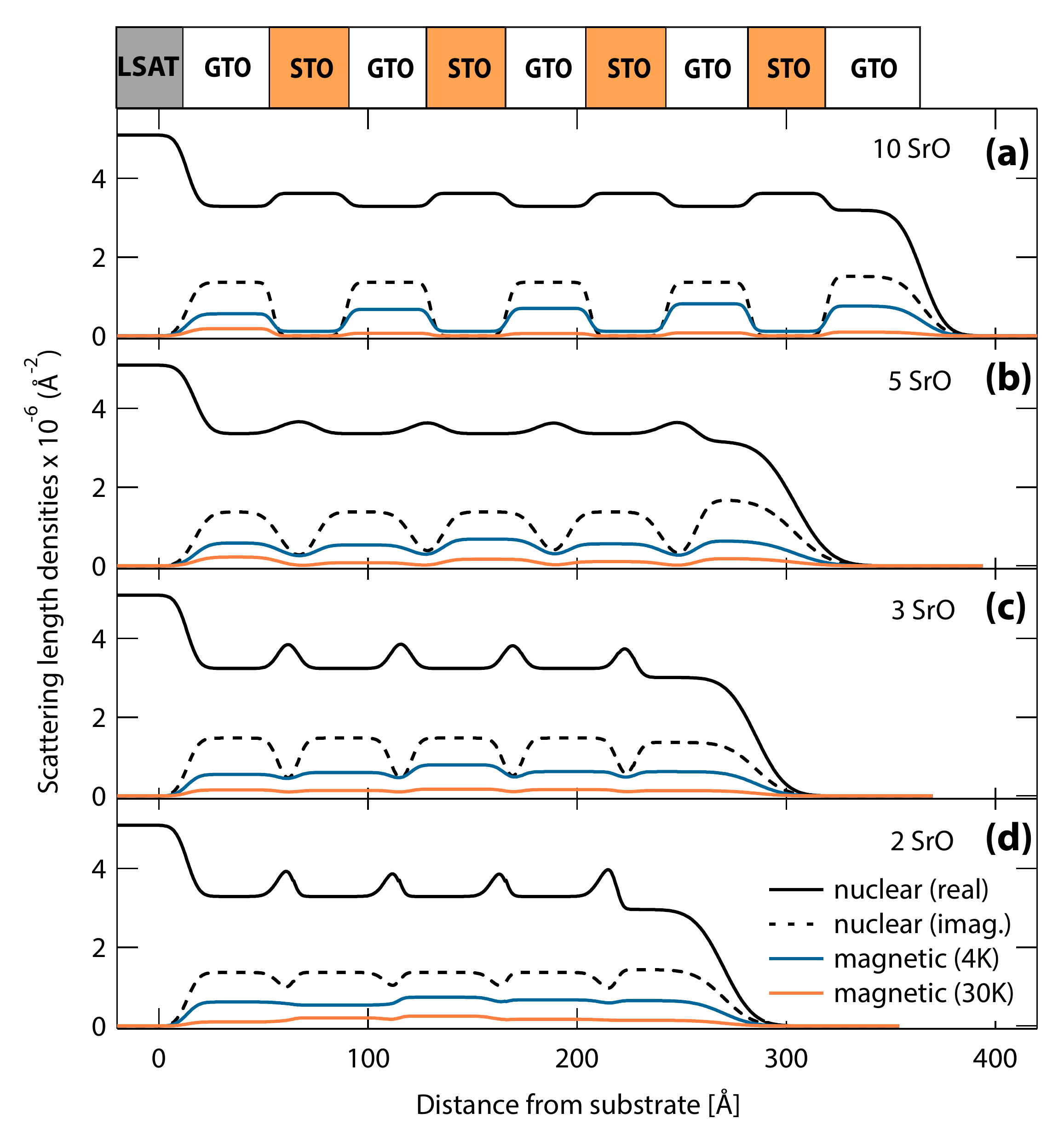}
\caption{Nuclear and magnetic depth profiles of GdTiO$_3$/SrTiO$_3$ superlattice films. SLDs as a function of film depth corresponding to the structural and magnetic models of PNR data for superlattice films with (a) 10 SrO, (b) 5 SrO, (c) 3 SrO, and (d) 2 SrO layers thick SrTiO$_3$ wells.  Solid and dashed black lines correspond to the real and imaginary components of nuclear scattering density profile, while the blue and orange lines correspond to the magnetization fits at $T = 4$ K and 30 K.}  
\end{figure}

The magnetic contrast between neighboring layers at 4 K notably does not follow the expected contrast between ferrimagnetic and nonmagnetic layers.  Instead, a finite magnetization persists across the SrTiO$_3$ wells for the 2 SrO, 3 SrO, and 5 SrO samples. To better demonstrate this, the magnetic components of the total scattering profiles are isolated by plotting the spin asymmetries, ($R^{++} - R^{--}$)/($R^{++} + R^{--}$) in Fig. 3.  Here, the low-$q$ portion of the asymmetry is dominated by the ferrimagnetism of GdTiO$_3$ spacers comprising the bulk of the sample.  At higher $q$ values, the scattering is more sensitive to magnetism associated with the SrTiO$_3$ quantum wells, particularly at a thickness-dependent Bragg position of the bilayer repeat.  Modeling the combination of these two extremes allows for the magnetic contrast between the layers to be directly refined.  

\begin{figure}
\includegraphics[scale=0.365]{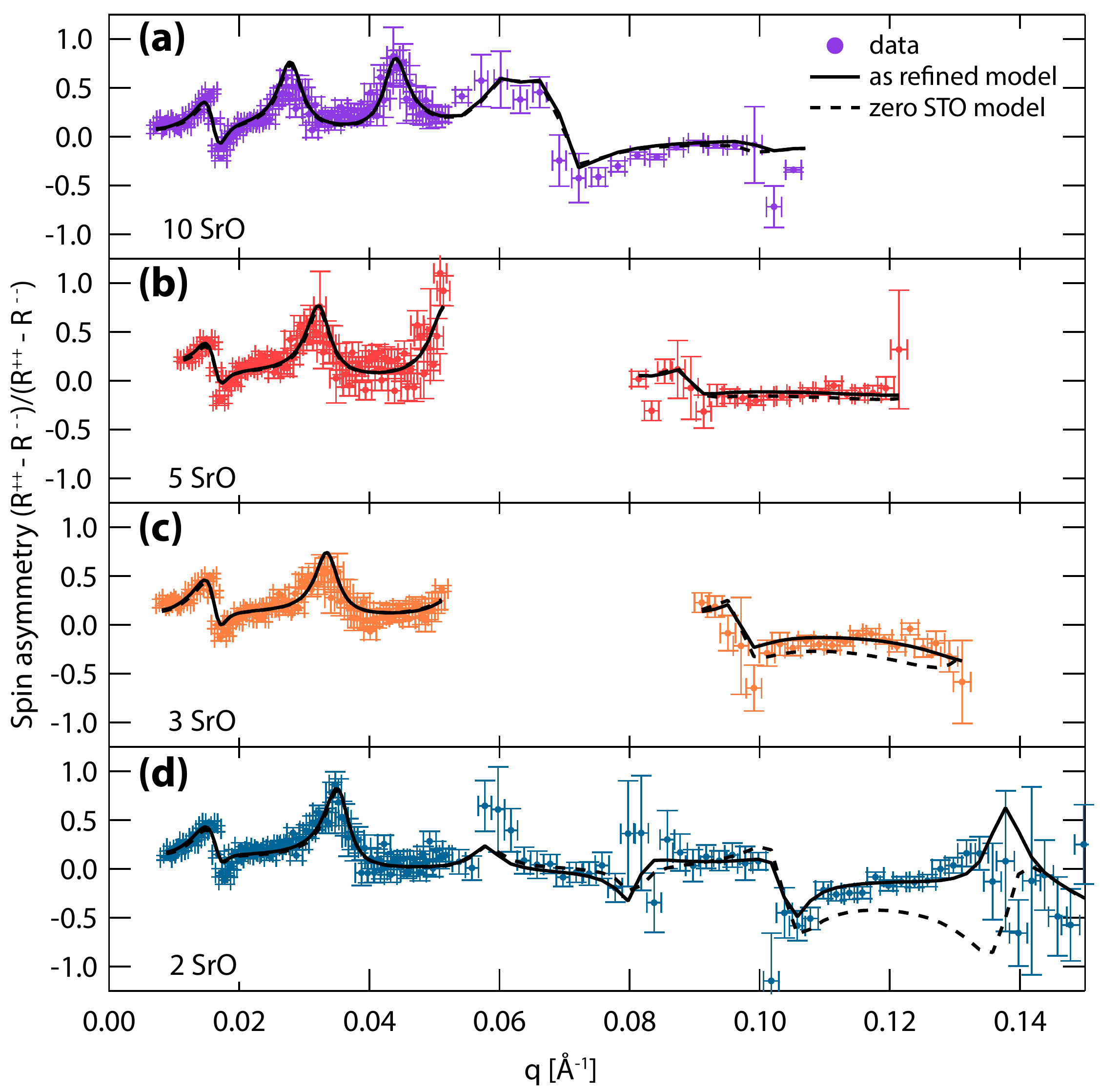}
\caption{Spin asymmetry at T = 4 K in GdTiO$_3$/SrTiO$_3$ superlattice films with (a) 10 SrO, (b) 5 SrO, (c) 3 SrO, and (d) 2 SrO layers thick SrTiO$_3$ wells.   Solid lines represent the refined models to the data, and the dashed lines represent a model constraining zero magnetization contribution from SrTiO$_3$.}  
\end{figure}

A qualitative sense of magnetism inside the thinnest 2 SrO wells is apparent via inspection of Figs. 1 (d), 2 (d), and 3 (d).  Fig. 1 (d) illustrates the Bragg peak and the corresponding $R^{++}$ and $R^{--}$ cross sections associated with the bilayer repeat at $q \approx 0.12 \AA^{-1}$.  The model profiles corresponding to these reflectivity curves plotted in Fig. 2 (d) show sharp contrast between the nuclear SLDs; however nearly negligible contrast is apparent within the magnetizations between layers.  In order to account for this diminished magnetic contrast, the presence of magnetism within the SrTiO$_3$ wells is illustrated in Fig. 3 (d).  Here a model of spin asymmetry allowing magnetized SrTiO$_3$ wells is compared with one forcing the magnetization contribution from SrTiO$_3$ to zero in the well center.  The freely refined model, placing finite magnetization in the SrTiO$_3$ wells, matches the data substantially better in the high $q$ limit where sensitivity to SrTiO$_3$ is maximal.  Stated in other words, spin asymmetry values near zero in the region of the Bragg peak necessitate a model with magnetism in the SrTiO$_3$ wells in order to produce the muted magnetic scattering contrast in the data. 

The average magnetization values in each superlattice (collected at the layer center values in model profiles) are plotted as a function of temperature for the four buried GdTiO$_3$ and SrTiO$_3$ layers in Figs. 4 (a) and (b), respectively.  Looking first at the spacer GdTiO$_3$ layers, ordered moment values show a temperature dependence tracking that of the ferrimagnetic order parameter observed in bulk crystals and relaxed films \cite{31,32,35}.  The small amount of scatter in the data arises from ambiguities in the absolute normalization of the reflectivity data \cite{Supplemental}, and taken as an average, the moments observed in the GdTiO$_3$ layers are 1.42 $\pm$ 0.20 $\mu_{B}/$f.u. at 4 K and 0.33 $\pm$ 0.11 $\mu_{B}/$f.u. at 30 K (f.u. = formula unit).  In order to confirm that the GdTiO$_3$ magnetization is independent of the well thicknesses, a separate film comprised of only a single 5 nm layer of GdTiO$_3$ was measured under identical conditions (i.e. $\mu_0 H = 0.7$ T FC).  The magnetization was refined to be 1.43 $\pm$ 0.13 and 0.24 $\pm$ 0.19 $\mu_B/$f.u. at 4 K and 30 K, respectively---within error of the superlattice values \cite{Supplemental}. The agreement between the magnetic properties of the isolated thin film and GdTiO$_3$ spacer layers confirms that the GdTiO$_3$ spacer layers are thick enough to decouple from the SrTiO$_3$ quantum wells.  This allows for added confidence in isolating the evolution of SrTiO$_3$ magnetism under varying well thickness.

\begin{figure}
\includegraphics[scale=0.4]{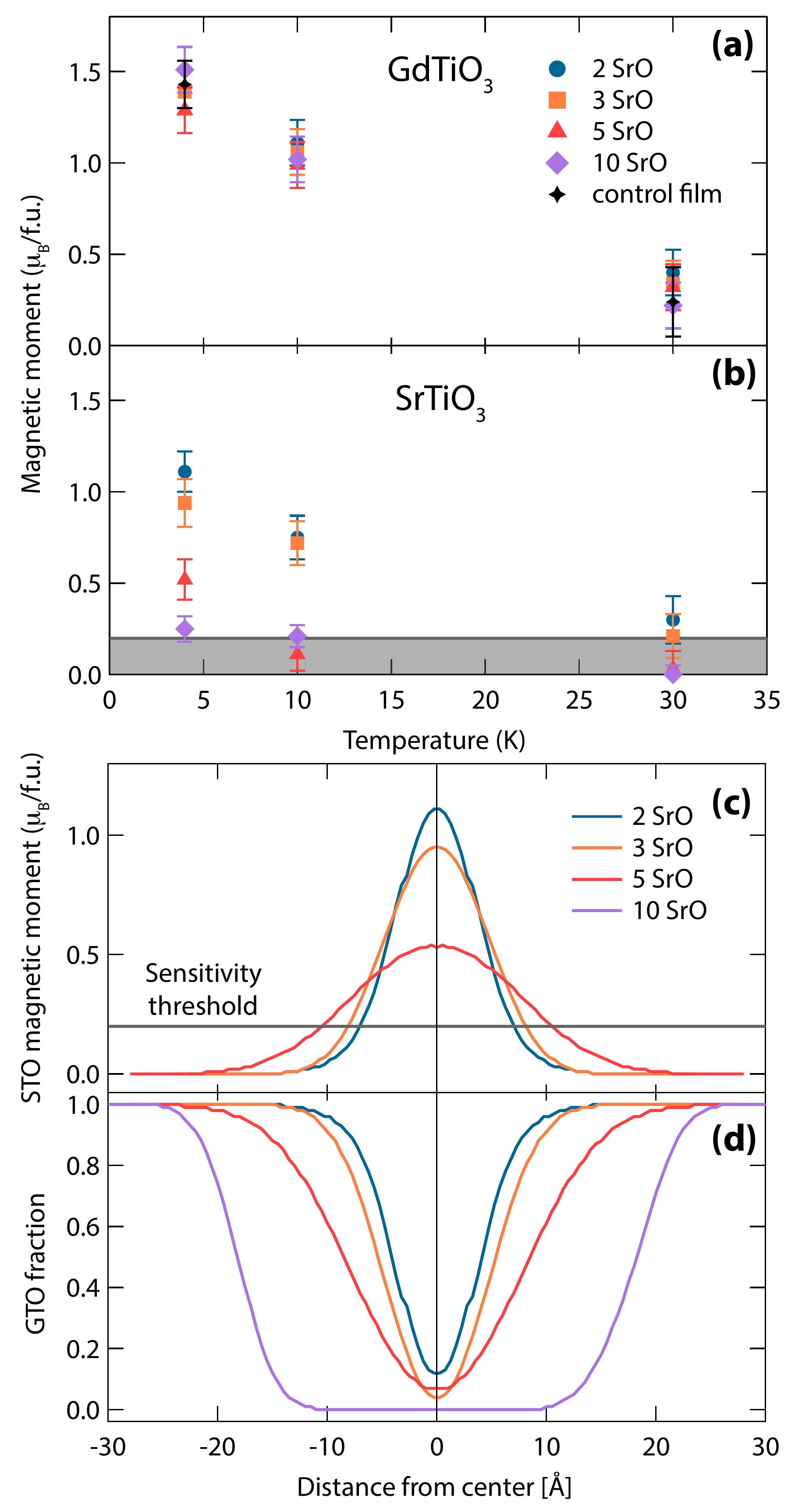}
\caption{(a) Magnetic moment per formula unit in GdTiO$_3$ layers plotted as a function of temperature for both superlattice films and the reference GdTiO$_3$ film as described in the text. (b) Magnetic moment per formula unit observed within SrTiO$_3$ layers plotted as a function of temperature in superlattice films.  Shaded region denotes approximate experimental sensitivity to SrTiO$_3$ moments. (c) Effective magnetization profiles for quantum wells after removing convolved GdTiO$_3$ contributions. (d) The relative fraction of GdTiO$_3$ convolved within the quantum wells as a function of distance from the well center.}  
\end{figure}

The magnetization values inherent to the SrTiO$_3$ layers are plotted in Fig. 4 (b).  For the three thinnest wells (2 SrO, 3 SrO, and 5 SrO), SrTiO$_3$ layers exhibit a finite magnetization whose temperature dependence seemingly tracks that of the polarizing GdTiO$_3$ spacer layers.  The saturated (4 K) moments in the SrTiO$_3$ layers increase as the well thicknesses are decreased and the electron gas at the interfaces is further confined; eventually reaching a peak value 1.11 $\pm$ 0.11 $\mu_B/$f.u. in the center of the 2 SrO wells.  This value is within error of the 1 $\mu_B/$Ti naively expected for fully polarized $S = 1/2$ Ti$^{3+}$ moments.  In contrast, the thickest 10 SrO sample refines to show a nearly vanishing SrTiO$_3$ magnetic moment within resolution (0.25 $\mu_B/$f.u. at 4 K).  Here the experimental sensitivity is effectively constrained by the uncertainty in the magnetization of the GdTiO$_3$ layers at this temperature. This is illustrated in Fig. 3(a) where models of the spin asymmetry containing magnetic versus nonmagnetic SrTiO$_3$ layers are identical.    

In order to parameterize the magnetism native to the quantum wells, the contribution of GdTiO$_3$ moments to the apparent magnetization of the SrTiO$_3$ wells can be largely accounted for and removed \cite{20}.  Specifically, the concentration of GdTiO$_3$ apparent within the wells can be calculated from the real parts of the nuclear SLD profiles by interpolating between pure GdTiO$_3$ and pure SrTiO$_3$.  This average convolution between layers is plotted in Fig. 4 (d) and represents the fraction of GdTiO$_3$ convolved into the SrTiO$_3$ layers as a function of displacement from the center of the wells.  In the thickest 10 SrO wells, the GdTiO$_3$ fraction drops to zero throughout the majority of the well, whereas in the thinnest 2 SrO wells the apparent roughness mixes in a substantial fraction of GdTiO$_3$ close to the well center.  This effective profile of GdTiO$_3$ within the wells can then be multiplied by the magnetization inherent to these spacer layers, yielding a maximum magnetic contribution from GdTiO$_3$ throughout the depth of the well.  The GdTiO$_3$ contribution is then subtracted from the total refined SrTiO$_3$ magnetization profile (c.f. Fig. 2), and the result is plotted in Fig. 4 (c).  This subtracted profile gives an average sense of how much of the refined moment is attributable to electrons induced within the well by the polar discontinuities.

The intrinsic ferromagnetism within the SrTiO$_3$ layers necessarily originates via the high-density two-dimensional electron gas (2DEG) induced at the interfaces.  Half of an electron per area unit cell is contributed to each well via the top and the bottom interfaces, yielding a total of one electron per well \cite{40}.  Therefore the integrated polarized moment in each well should be a constant value of 1 $\mu_B$ regardless of the well thickness. From our models, we calculate total integrated moments in the wells to be 2.75 $\mu_B$, 2.83 $\mu_B$, and 2.60 $\mu_B$/well, for the 2, 3, and 5 SrO samples, respectively.  Although these values are consistent with a picture of a constant integral moment, their magnitude likely reflects an inherent overestimation born by modeling the magnetization density as peaked in the center of the SrTiO$_3$ wells \cite{Supplemental}.  

While \textit{f-d} hybridization effects may play a role in polarizing some fraction of electrons directly at the interfaces, the majority of induced electrons are known to substantially delocalize into the volume of the SrTiO$_3$ wells and suggest a more extended exchange mechanism \cite{7}.  The density of interface-induced carriers nominally decays by $50\%$ over approximately 1 nm into the bulk of SrTiO$_3$ \cite{17}, and the average 1.8 nm thick wells of the 5 SrO sample are consistent with a threshold where the overlap between interface states becomes appreciable.  Moving substantially above this thickness corresponds to distances where the extended 2DEGs stabilized at each polar interface no longer sufficiently overlap and support the continuation of the exchange field across the well. We stress here however that our PNR measurements are unable to comment on presence of magnetic texture within the wells themselves, rather, in the thin well limit, the resolution of our data only permits effective models of uniformly magnetized wells.  Despite this, the disappearance of SrTiO$_3$ magnetism with increasing well thickness connects ferromagnetic spin correlations in the SrTiO$_3$ wells with the local order parameter destabilized at the quantum critical point in this system \cite{18}.    

The temperature dependence of the ordered moments within the wells tracks that of the ferrimagnetism within GdTiO$_3$ spacing layers, suggesting that the molecular field of neighboring GdTiO$_3$ polarizes moments within the wells.  The induced phase is therefore distinct from the hysteretic response identified in prior magnetoresistance measurements with a lower characteristic temperature (T$_c \approx 5$ K).  Either a nontrivial field dependence of the order induced within the wells or an alternative order parameter, such as orbital order, should be invoked to explain this low temperature state.  Rather, our key finding is a striking realization of interface-induced magnetic polarization across nominally nonmagnetic SrTiO$_3$ quantum wells nearly 2 nm thick embedded within a Mott insulating GdTiO$_3$ matrix.     \nocite{23,22,24,25}

\acknowledgments{
S.W., R. N. and S.S. acknowledge support under ARO award number W911NF1410379.   R.N. was supported in part by the National Science Foundation Graduate Research Fellowship under Grant No. 1144085. }

\bibliography{BibTeX_GTOSTOPNR}

\end{document}